\documentclass[
    ,final            
  ]
  {aipproc}

\layoutstyle{8x11double}
\usepackage{amssymb}   
\usepackage{amsmath}
\usepackage{amsfonts}
\usepackage{amssymb}
\begin{document}
\newcommand{\beqn}{\begin{eqnarray}}
\newcommand{\eeqn}{\end{eqnarray}}
\newcommand{\be}{\begin{equation}}
\newcommand{\ee}{\end{equation}}
\newcommand{\non}{\nonumber \\}
\def \ETmiss{${\not\!\!{E_T}}~$}
\def \missET{${\not\!\!{E_T}}$}

\newcommand{\nuni}{nonuniversalities~}
\def \cha{\widetilde{\chi}^{\pm}_1}
\def \chb{\widetilde{\chi}^{\pm}_2}
\def \na{\widetilde{\chi}^{0}_1}
\def \nb{\widetilde{\chi}^{0}_2}
\def \nc{\widetilde{\chi}^{0}_3}
\def \nd{\widetilde{\chi}^{0}_4}
\def \g{\widetilde{g}}
\def \ql{\widetilde{q}_L}
\def \qr{\widetilde{q}_R}
\def \dl{\widetilde{d}_L}
\def \dr{\widetilde{d}_R}
\def \ul{\widetilde{u}_L}
\def \ur{\widetilde{u}_R}
\def \ccl{\widetilde{c}_L}
\def \ccr{\widetilde{c}_R}
\def \ssl{\widetilde{s}_L}
\def \ssr{\widetilde{s}_R}
\def \ta{\widetilde{t}_1}
\def \tb{\widetilde{t}_2}
\def \ba{\widetilde{b}_1}
\def \bb{\widetilde{b}_2}
\def \sta{\widetilde{\tau}_1}
\def \stb{\widetilde{\tau}_2}
\def \smr{\widetilde{\mu}_R}
\def \ser{\widetilde{e}_R}
\def \sml{\widetilde{\mu}_L}
\def \sel{\widetilde{e}_L}
\def \slr{\widetilde{l}_R}
\def \sll{\widetilde{l}_L}
\def \snl{\widetilde{\nu}_{\tau}}
\def \snm{\widetilde{\nu}_{\mu}}
\def \sne{\widetilde{\nu}_{e}}
\def \hc{H^{\pm}}
\def \lra{\longrightarrow}

\title{
\begin{flushright}
{\small CERN-PH-TH/2008-143}
\end{flushright}
Recent Developments in Supersymmetric \\and Hidden Sector Dark Matter}

\classification{95.35.+d,04.65.+e,11.30.pb}
\keywords      {SUSY Dark Matter, mSUGRA, NUSUGRA, LHC Signatures, milli-charged Dark Matter}

\author{Daniel Feldman}{
  address={Department of Physics, Northeastern University, Boston, MA 02115, USA.}
}

\author{Zuowei Liu}{
  address={Department of Physics, Northeastern University, Boston, MA 02115, USA.}
}

\author{Pran Nath}{
address={Department of Physics, Northeastern University, Boston, MA 02115, USA.},
  altaddress={TH Division, PH Department, CERN, CH-1211 Geneva 23, Switzerland.}
}

\begin{abstract}
New results which correlate SUSY dark matter with LHC signals are presented,
and a brief review of recent developments in supersymmetric and hidden sector dark matter
is given.
It is shown that the direct detection of dark matter is very sensitive to the hierarchical
SUSY sparticle spectrum and the spectrum is very useful in distinguishing models.
It is shown that the prospects of the discovery of neutralino dark matter are
very bright on the "Chargino Wall" due to a copious number of model points on
the Wall, where the NLSP is the Chargino, and the spin independent neutralino-proton cross section
is maintained at high values in the $10^{-44}$cm$^{2}$ range for neutralino masses up
to $\sim 850$ GeV .  It is also shown that the direct detection of dark matter along with
lepton plus jet signatures and missing energy provide dual, and often complementary,
probes of supersymmetry. Finally, we discuss an out of the box  possibility for dark matter,
which includes dark matter from the hidden sector, which could either consist of
extra weakly interacting dark matter (a Stino XWIMP), or milli-charged dark matter arising
from the Stueckelberg extensions of the MSSM or the SM.
\end{abstract}

\maketitle

\section{Introduction}
We will discuss here three topics. These are: (1)
Neutralino dark matter from the perspective of a
sparticle landscape\cite{Feldman:2007zn},
and sparticle mass hierarchies; (2)
The dual probe of supersymmetry with dark matter detection, and
with leptonic and jet signatures and missing energy
from sparticle production at the LHC\cite{Feldman:2008hs};  (3) An out of the
box possibility of extra weakly interacting dark matter arising from the
hidden sector\cite{Kors:2004ri,Feldman:2006wd} and mill-charged\cite{holdom,Kors:2004dx}
 dark matter\cite{Cheung:2007ut,Feldman:2007wj}.

 We begin with supersymmetry
 which is an attractive symmetry for the construction of fundamental
interactions in four dimensions\cite{Wess:1973kz}. For a variety of reasons supersymmetry
must be made local\cite{Nath:1975nj,Freedman:1976xh} which leads to what one calls
supergravity. Further support for supergravity comes from the fact that it is the
field point limit of string theory which is a candidate theory of quantum gravity.
 The above  provides the rationale for utilizing  $N=1$ supergravity as
a natural framework for model building. In this class fall the sugra, string and  D brane
models.  These are all high scale models which differ  among other things, by the
nature of soft  SUSY breaking.  Soft  breaking can be classified broadly as arising from
 gravity mediation\cite{chams},\cite{barbi},\cite{hlw},\cite{Nilles:1983ge}, from gauge mediation\cite{Giudice:1998bp},
as well as other possibilities such as from anomaly mediation etc. In this analysis we will focus
on the gravity mediation of soft breaking. The  minimal supergravity models are characterized
by the parameter space in the soft sector at the GUT scale
 consisting of the four parameters  $(m_0,  m_{1/2}, A_0, B_0)$  where ($m_0$, $m_{1/2}$) are the universal (scalar,gaugino) masses,
$A_0$ is the universal trilinear coupling, and $B_0$ is the parameter which appears as $B\mu_0 H_1H_2$, where
$\mu$ is the co-efficient of the bilinear Higgs  term in the superpotential which appears in the
form $\mu H_1H_2$.
After  radiative electroweak symmetry breaking (REWSB) one determines $\mu$ except for it sign, and trades
$B_0$ for $\tan\beta$ which is defined to be the ratio of two Higgs VEVs, i.e.,
$\tan\beta=\langle H_2 \rangle/\langle H_1 \rangle$, where $(H_2,H_1)$ are responsible for mass generation of (u quarks, d quarks and leptons).
Thus the parameter space
of the model after REWSB is spanned by $(m_0,  m_{1/2}, A_0, \tan\beta, {\rm sign} (\mu))$ (see, e.g., \cite{an}).
The mSUGRA model is precisely defined to be a model  with the parameter space specified above.

We note that supergravity models provide a broad framework for model building. Thus one can both reduce the
parameter space of the model by additional constraints such as  in no-scale models or by putting further constraints on
the mSUGRA parameter space, or enlarge the parameter space by inclusion of
non-universalities.
 Models with enlarged parameter space include
sugra models with non-universalities, heterotic string and  D brane models with large volume compactifications,
 and many other scenarios\cite{NU,Kors:2004hz}. We note that irrespective of the details of the models, all
 models of this sort fall in the general class where SUSY is broken by gravity mediation.
 Some generic non-universalities  in sugra models (one can label such models as NUSUGRA)
 can be discussed by the inclusion of non-universalities in the
Higgs sector (NUH), in the 3rd generation sector (NUq3), and in the gaugino sector (NUG).
Thus, for example, in the NUH case one can include  \nuni at the GUT scale so that
(i) NUH: $M_{H_u}
= m_0(1+\delta_{H_u})$,  $M_{H_d} = m_0(1+\delta_{H_d})$.
Similarly for the third generation one can include \nuni such that  (ii) NUq3:
$M_{q3}=m_0(1+\delta_{q3})$,  $M_{u3,d3}=m_0(1+\delta_{tbR})$,  and finally  for
the gaugino sector one may include \nuni such that
(iii) NUG: $M_{1}=m_{1/2}$, $M_{2}=m_{1/2}(1+\delta_{M_2})$,
$M_{3}=m_{1/2}(1+\delta_{M_3})$. In all cases the $\delta$'s parameterize the \nuni and one may
take their ranges to lie in some reasonable interval such as $-0.9\leq\delta\leq
1$.\\

In supergravity models the lightest neutralino turns out to be the lightest supersymmetric
particle (LSP)  over a large region of the parameter space, and is thus a candidate for
dark matter with R parity.
For neutralino dark matter the satisfaction of the WMAP constraints\cite{Spergel:2006hy}
$.0855 < \Omega_{cdm} h^2 < .1189 ~~(2\sigma)$
 is achieved typically in three broad regions. These include the
co-annihilation regions, the Hyperbolic Branch/Focus Point (HB/FP)  region\cite{hb/fp,Lahanas:2003bh} and
pole regions.
The   co-annihilation regions contain stau co-annihilation \cite{Ellis:1998kh}, stop co-annihilations
etc. The relic density analysis allows  a region of the parameter space where the
CP odd Higgs is light and  where WMAP constraints
are also  satisfied. Recently light Higgses in the context of neutralino dark
matter have been discussed in \cite{Drees,Carena:2006nv,Ellis:2007ss,Feldman:2007fq,Barenboim:2007sk,LH1,LH2}.
Based on restricted analyses  it is often stated that
only small slivers of the mSUGRA parameter space remain consistent with WMAP.
However, this conclusion is  erroneous since a large part of the parameter space opens
up when $A_0$ and $\tan\beta$ are fully explored\cite{modern3,modern4,Feldman:2007zn}.
There is an enormous literature on the analyses of SUSY dark matter. A small sample
can  be  found in \cite{modern,modern1,Drees:2008bv,Bottino:2008xc}.

\section{HIERARCHICAL MASS PATTERNS }
An approach which has proved useful in the analysis of dark matter
and in  correlating it with the LHC physics is in terms of sparticle mass patterns
 \cite{Feldman:2007zn,Feldman:2007fq,Feldman:2008hs}.
As there are 32 sparticle masses in MSSM (including Higgses in this definition), then using sum rules
one has upwards of $10^{25}$ mass hierarchies. If one focusses on the first four lightest
sparticles this number reduces to about $10^4$.  It reduces further, and quite drastically, in
well motivated models such as mSUGRA, NUSUGRA,
and in string and D brane models when one imposes the accelerator and WMAP constraints,
and the constraints of REWSB.  For  the case of mSUGRA, with $\mu>0$ one finds that 16 patterns survive
(labeled mSP1-mSP16 and can be decomposed more simply in terms of the NLSP):\\
 \noindent
 {\small
 {\bf Chargino Patterns } ($\mu>0$)\\
{\bf mSP1}:  ~ $\na$   $<$ $\cha$  $<$ $\nb$   $<$ $\nc$,
{\bf mSP2}: ~  $\na$   $<$ $\cha$  $<$ $\nb$   $<$ $A/H$,
{\bf mSP3}: ~ $\na$   $<$ $\cha$  $<$ $\nb$ $<$ $\sta$,
{\bf mSP4}: ~ $\na$   $<$ $\cha$ $<$ $\nb$   $<$ $\g$,\\
 {\bf Stau Patterns } ($\mu>0$)\\
{\bf mSP5}: ~ $\na$ $<$ $\sta$  $<$ $\slr$  $<$ $\snl$,
{\bf mSP6}: ~  $\na$   $<$ $\sta$  $<$ $\cha$  $<$ $\nb$,
{\bf mSP7}:  ~  $\na$   $<$ $\sta$  $<$ $\slr$  $<$ $\cha$, ~
{\bf mSP8}: ~  $\na$ $<$ $\sta$  $<$ $A\sim H$,\\
{\bf mSP9}: ~ $\na$   $<$ $\sta$  $<$ $\slr$ $<$ $A/H$,
{\bf mSP10}: ~ $\na$   $<$ $\sta$ $<$ $\ta$ $<$ $\slr$ \\
{\bf Stop Patterns}  ($\mu>0$)\\
{\bf mSP11}: ~ $\na$ $<$ $\ta$ $<$ $\cha$  $<$ $\nb$,
{\bf mSP12}: ~ $\na$ $<$ $\ta$   $<$ $\sta$ $<$ $\cha$,
{\bf mSP13}:~  $\na$   $<$ $\ta$ $<$ $\sta$ $<$ $\slr$ \\
{\bf Higgs Patterns} ($\mu>0$)\\
{\bf mSP14}:~ $\na$   $<$  $A\sim H$ $<$ $\hc$,
{\bf mSP15}:~ $\na$   $<$ $ A\sim H$ $<$ $\cha$,
{\bf mSP16}: ~ $\na$   $<$ $A\sim H$ $<$$\sta$ .\\}

The notation mSP stands for minimal SUGRA Pattern. For the case $\mu<0$ one  finds more
Stau and Stop Patterns and additionally a new type appears which is
the neutralino pattern : \\
{\small
{\bf Stau Patterns}  ($\mu<0$)\\
{\bf mSP17}:~ $\na$   $<$ $\sta$ $<$ $\nb$ $<$ $\cha$,
{\bf mSP18}:~ $\na$   $<$ $\sta$  $<$ $\slr$  $<$ $\ta$,
{\bf mSP19}:~  $\na$ $<$ $\sta$ $<$ $\ta$   $<$ $\cha$\\
{\bf Stop Patterns}  ($\mu<0$)\\
{\bf mSP20}: ~ $\na$ $<$ $\ta$   $<$ $\nb$   $<$ $\cha$,
{\bf mSP21}:~ $\na$   $<$ $\ta$   $<$ $\sta$  $<$ $\nb$\\
{\bf Neutralino Pattern}  ($\mu<0$)\\
{\bf mSP22}:  $\na$   $<$ $\nb$   $<$ $\cha$  $<$ $\g$ . \\}

\begin{figure*}
  \includegraphics[height= 6cm , width = 8cm]{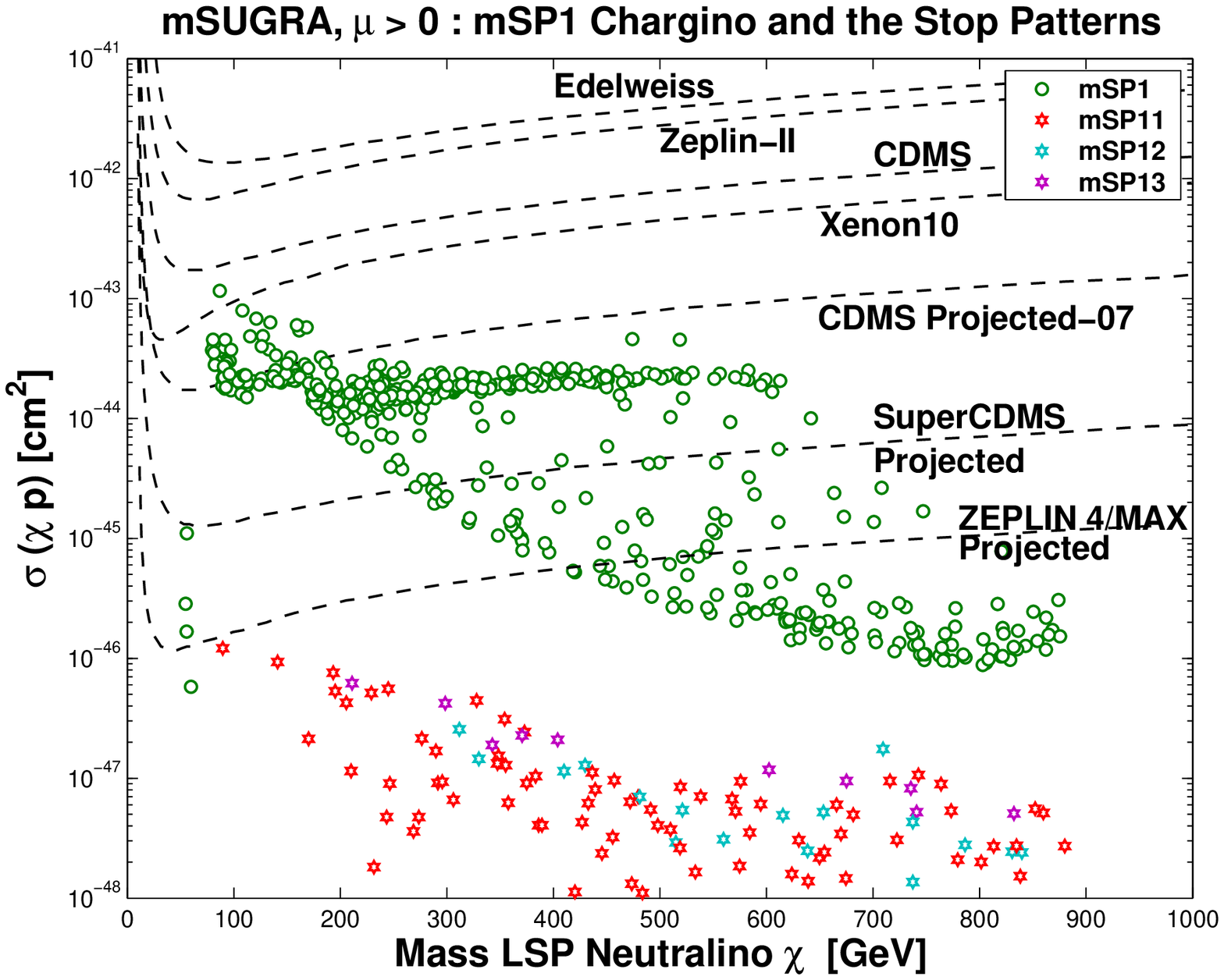}
 \includegraphics[height= 6cm , width = 8cm]{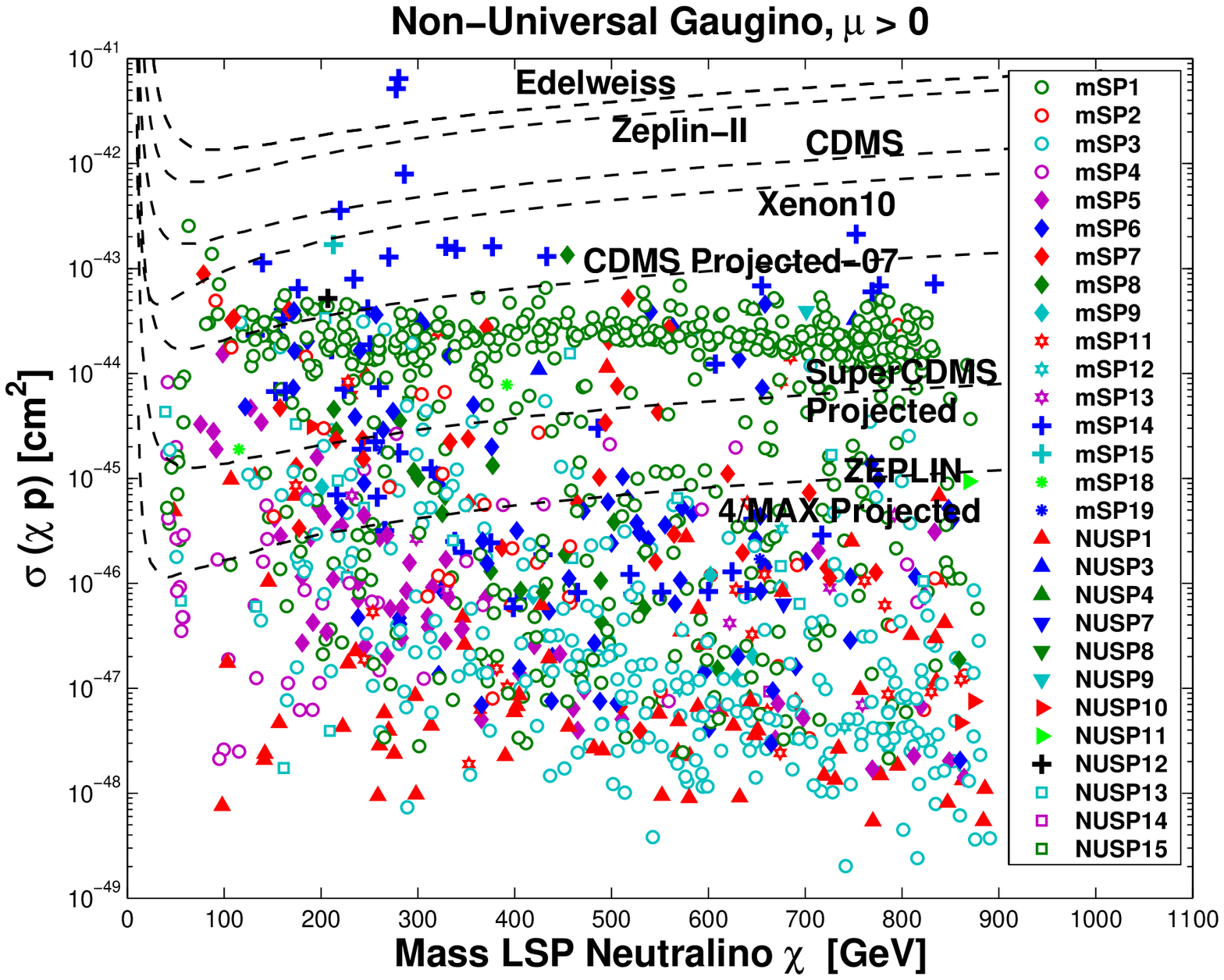}
  \caption{An exhibition of the Chargino Wall for the mSUGRA ($\mu>0$) case (left panel) and for the
  NUG   case  with \nuni in the gaugino sector (right panel). The analysis shows that Chargino Wall
  consisting of a copious number of mSP1 points sustains  an LSP up to 850 GeV
  with spin independent cross sections at the $10^{-44}$ cm$^2$ level, see \cite{Feldman:2007fq}.
  \label{dmddfig}}
\end{figure*}
We note that only 6 of the 22 patterns listed above are sampled in the
Snowmass\cite{Allanach:2002nj},  in the PostWMAP3\cite{Battaglia:2003ab} and in the
 CMS LM and HM benchmarks.
Since it is imperative that one sample all the patterns,  benchmarks for
the 22 patterns have been given recently in \cite{Feldman:2008hs}.
As an example,  an application  using mSP4 is given in \cite{Gounaris:2008pa}.
With the inclusion of \nuni in the soft breaking sector
for the cases of NUH, NUq3, and NUG  15 more mass patterns emerge which may be
labeled as NUSP1-NUSP15  (see \cite{Feldman:2008hs}).

It turns out  that the direct detection  of
dark matter (for early works see e.g. \cite{Ellis:1987sh,Drees:1992am}, \cite{Falk:1998xj,Chattopadhyay:1998wb})
produces a strong  dispersion in the patterns \cite{Feldman:2007fq}. An example of this is given in
Fig.(\ref{dmddfig}) where one finds  a large dispersion between the Chargino
Patterns and the Stop Patterns  (more examples of model discrimination with dark matter connected to LHC signatures may be
found in \cite{Feldman:2007fq} and also in \cite{nelson}).  Another interesting phenomenon is the appearance of
the Chargino Wall in mSP1 which runs horizontally up to  $\sim$ 650 GeV in the neutralino mass for
mSUGRA and up to $\sim$ 850 GeV in the neutralino mass for the NUG model under naturalness assumptions.
Here one finds that the spin independent cross section is maintained at
$\sim (2-5)\times 10^{-8}$ pb level  over the entire range of neutralino mass enhancing the
prospects for the discovery of dark matter on the Wall in upgraded dark matter experiments.
We add that while a larger Higgsino content is known to give rise to strong SI cross sections \cite{dmdd} the finding that the Wall is composed essentially entirely of mSP1 points in sugra
models\cite{Feldman:2007fq} is an entirely new result  which also  has important
implications  for LHC studies.
 In addition to the neutralino, there are other alternatives dark matter candidates such as
the gravitino in sugra models,  the least massive KK particle (LKP) as, e.g.  in UED models; a massive spin 1 in
Little Higgs Models, Dirac neutrinos,
dark matter from the hidden sector, and several other interesting possibilities.
A recent work has observed\cite{Barger:2008qd}
 that a comparison of spin dependent vs spin independent
scattering cross sections can be used to distinguish some of the models listed above.

\subsection{
Dual probes of SUSY with dark matter detection
and + leptons and jets + \missET }
\begin{figure}[h]
  \includegraphics[height=.25\textheight]{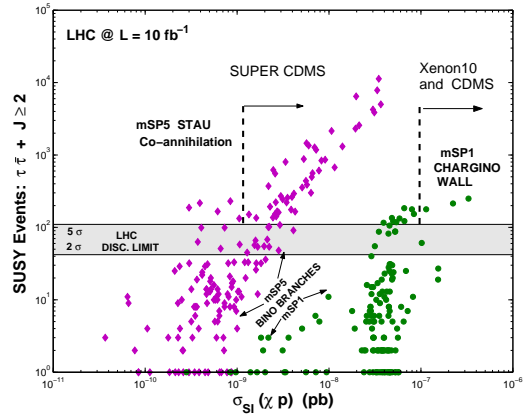}
  \caption{ An exhibition of the dual probes of SUSY by direct detection experiments and
   by lepton, jet and missing energy signals at the LHC. The analysis above focuses on the
   Chargino Pattern mSP1 and the Stau Pattern mSP5 for mSUGRA ($\mu > 0$), and here we use the new module of
   \cite{Belanger:2008sj} with $\sim 330$ simulated SUSY models employing Pythia \cite{Sjostrand:2006za} and PGS4 \cite{pgs}
   ranging over the soft parameter space with SuSpect \cite{SUSPECT} for $m_0 < 4$ TeV, $m_{1/2} < 2 $ TeV, $|A_0/m_0| < 10$
   and for $\tan \beta \in (1,60)$.
 Relic density, mass limits and FCNC constraints and cuts for SM backgrounds are as in \cite{Feldman:2008hs}.}
   \label{dual}
\end{figure}

It is important to pursue correlated studies of
experimentally constrained dark matter \cite{wim} (see also \cite{Bednyakov:2008gv})  with  signatures at the Large Hadron Collider.
Some recent analyses of LHC signature spaces have been studied in
\cite{Feldman:2007zn,Feldman:2007fq,Feldman:2008hs,kks,Conlon:2007xv,arnowitt,mmt,bps,Lafaye:2007vs,othertalks1,othertalks2}.
One finds that dark matter detection is in some ways complementary to LHC in its
probe of the SUSY parameter space.  That is, dark matter direct detection can probe some parts of the parameter
space which may be hard to reach with low luminosity at the LHC. One such example is given
 in Fig.(\ref{dual})
where one finds that  a much larger part of the parameter space of
chargino patterns can be explored with Super CDMS  (which covers the whole Wall)
than with 10 fb$^{-1}$ of integrated LHC
luminosity in the OS $2\tau$ channel. The plot shows remarkable separation between
the stau co-annihilation region and the hyperbolic branch.

\section{Hidden sector Dark matter }
We discuss now an out of the box possibility for dark matter. An interesting possibility arises
in that dark matter can originate from a  hidden sector. In sugra unified models and in string
and in brane models a hidden sector exists which contains fields which are singlets of the
Standard Model gauge group. Thus it is interesting to investigate if the hidden sector can
provide us with the relevant candidate for dark matter which produces relic density within
the WMAP bounds.
Suppose there is dark matter whose interactions with quarks and leptons
 are weaker than weak, or extra-weak.
How can such dark matter arise?
Such extra-weak dark matter can arise when one has two sectors: a physical sector
where MSSM fields reside and a hidden sector. The hidden sector fields do not carry MSSM
quantum numbers and the physical sector fields do not carry the quantum numbers of fields
in the hidden sector.  Thus the sectors do not have a direct communication.

If, however, one introduces
a connector sector which carries dual quantum numbers and interacts with the
physical sector fields as  well as with the hidden sector fields then the sectors can communicate
\cite{Feldman:2006wd}.
Further, spontaneous breaking in the connector sector would produce mixing effects in the
mass  matrices in the visible sector which can lead to detectable signals.
We give now an explicit demonstrations of the above.
We begin by
considering for the hidden sector just a $U(1)_X$ gauge multiplet .
  For the Connector Sector we consider the chiral fields $\phi^{\pm}$ with charges $\pm Q_X$ under $U(1)_X$ and charges $\pm Y_{\phi}$ under $U(1)_Y$.
  For technical reasons one needs to add  a  Fayet-Iliopoulos term
$\mathcal{L}_{FI}= \xi_X D_X +\xi_Y D_Y$.
Vacuum solutions for this model  give $\langle \phi^+ \rangle=0$, and $\langle \phi^- \rangle \neq 0$
and one has mixings involving  the visible sector, the hidden sector and the connector sector.
We discuss the implication of this mixing for dark matter.

After spontaneous breaking
 there are now six Majorana spinors
 $
 (\chi_{\phi^-}, \lambda_X; \lambda_Y, \lambda_3, \tilde h_1, \tilde h_2)
$
where $\chi_{\phi^-}$ is the spinor that arises from  $\phi^{-}$.  This leads to mass diagonal  states
$
(\xi_1^0, \xi_2^0); (\chi_1^0,\chi_2^0,\chi_3^0,\chi_4^0).
$
 The above scenario is actually realized in the Stueckelberg $U(1)_X$ extension of the MSSM \cite{Kors:2004ri,Kors:2005uz}.
 In the St extensions  there is a mixing that occurs between the two $U(1)$ factors in the theory,
 i.e., $U(1)_X$ and $U(1)_Y$ which arises from the following Lagrangian:
 \beqn
  \mathcal{L}_{St}(V,S,\bar S) = ({M_1 C + M_2 B + S +\bar S})^2|_{\theta \theta \bar \theta  \bar \theta}
 ,
 \eeqn
where $V=(C,B)$ are vector superfields and $S$ is a chiral superfield. In the vector field sector
this leads to in particular
the combination $(-1/2)(\partial_{\mu}\sigma + M_2 B_{\mu}
 + M_1 C_{\mu})^2$ where $B_{\mu}$ is the gauge field of $U(1)_Y$ and $C_{\mu}$ is the gauge
 field of $U(1)_X$, and $\sigma$ is the axion which gets absorbed in the unitary gauge.
 The parameter that produces mixings between the visible sector and the hidden sector is
 $\epsilon =M_2/M_1$, and an analysis based on precision electroweak data gives
 the constraint $\epsilon \lesssim .06$. Because of the smallness of $\epsilon$ the interactions
 of $\xi_1^0$ and $\xi_2^0$ with the visible sector quarks and leptons are extra  weak.
 Using the index 1 to denote the lighter of each type of Majorana we now have the following situation:
 either one has $m_{\xi_1^0}>m_{\chi_1^0}$ or one
 has  $m_{\xi_1^0}< m_{\chi_1^0}$. For the case when $m_{\xi_1^0}> m_{\chi_1^0}$,
 ${\chi_1^0}$ will still be the LSP and not much will change. However, for the case when
 $m_{\xi_1^0}< m_{\chi_1^0}$ it is ${\xi_1^0}\equiv {\xi^0}$  which is the LSP, and the LSP in this case
 will be extra weakly interacting. We will call this particle an XWIMP or Stino for obvious
 reasons.
 XWIMPS cannot annihilate in sufficient amounts by themselves to satisfy the
relic density constraints as mentioned already. However, they can do so via co-annihilation, i.e., via the processes
$\xi^0+\xi^0 \to X$,  $\xi^0+\chi^0\to X'$, and  $\chi^0 +\chi^0\to X''$ where each $X$ are
 (pairs of) Standard Model particle.
The  effective cross section for the annihilation of the extra-weakly interacting Stinos is then,
\beqn
\sigma_{eff} \sim \sigma_{\chi^0\chi^0} (\frac{Q}{1+Q})^2, ~~Q\sim  (1+\Delta)^{3/2} e^{-x_f\Delta},
\eeqn
where $\Delta =(m_{\chi^0}- m_{\xi^0})/m_{\xi^0}$,  $x_{f}= m_{\xi^0}/T_f$  and $T_f$ is
the  freeze out temperature.  For $x_f \Delta <<1$,  $Q\sim 1$
and one can produce enough co-annihilation to  efficiently annihilate the XWIMPs, and find their relic  density within the WMAP  range.
The above can be generalized to include other MSSM channels.

The second case of dark matter from the hidden sector that we consider is the case of
milli-charged dark matter. It has been known for some time \cite{holdom}
that
milli-charged matter arises from the kinetic mixing with two $U(1)$s
through a mixing term (defined here by $\delta$) generated by exchange of heavy fields. Such mixings can survive at low energy.
In the diagonal basis one gets two massless gauge bosons, one of which is the
ordinary photon ($A_{\mu}$) and the other
a (massless) paraphoton ($A_{\mu}'$).   In an appropriate basis the interactions  can be written
in the form
$  A \cdot (J +\delta  J^{\rm hid}) +A' \cdot J^{\rm hid}$. Here  the photon couples to the
hidden sector matter fields with a coupling
proportional to the small mixing $\delta$ which is generated by the exchange of heavy
fields, while the paraphoton does not couple with the visible sector and only couples to the hidden sector.
Dark matter in this model has been analyzed in \cite{Goldberg:1986nk}.
Milli-charged   matter also arises in Stueckelberg extensions of the
Standard Model\cite{Kors:2004dx,Kors:2005uz}
where two $U(1)$ gauge fields mix via mass mixing. Such models can arise from string
constructions \cite{Kiristis,Abel}. Some recent works involving St mass generation can be found in
\cite{KirCorAna,Burgess,Ahlers:2007qf}.

The St models can also sustain milli-charged dark matter\cite{Cheung:2007ut,Feldman:2007wj}.
The St models we will discuss here includes both mass and kinetic mixing via a Lagrangian of the form \cite{Feldman:2007wj}
\beqn
\mathcal{L}_{StKM} \supset -\frac{1} {4} [C_{\mu\nu} C^{\mu\nu}+ 2\delta  C_{\mu\nu} B^{\mu\nu}
+B_{\mu\nu} B^{\mu\nu}]\nonumber\\
-\frac{1}{2} (\partial_{\mu} \sigma + M C_{\mu} +\epsilon M B_{\mu})^2 +
J_{\mu} B^{\mu} + J_{\mu}^{\rm hid}C^{\mu}.
\eeqn Here one can have both mass mixing ($\epsilon$) and kinetic
mixing ($\delta$). In the diagonal basis there is only one massless
mode (normal photon) and the other vector boson modes are all massive. For
$Z',A_{\gamma}$ there are interactions  of the generic form
\beqn
\mathcal{L}^{Int}_{StKM} \sim f_1((\epsilon -\delta)J_{\mu} + f_2 J^{{\rm hid}}_{\mu}) Z^{'\mu} + f_3(J_{\mu}^{vis}- \epsilon J^{hid}_{\mu}) A_{\gamma}^{\mu}.\nonumber \eeqn
where $f_{1,2,3} = f_{1,2,3}(\epsilon,\delta)$ (see \cite{Feldman:2007wj} for the complete form).
The  constraints on $\epsilon$ and $\delta$ are gotten by fits to the
precision electroweak data, where one finds for example
$(\epsilon,\delta) = (.06,.03)$ can fit the data with the same
precision as does the SM. If there is hidden sector matter it would
carry milli-charge. An interesting possibility is that such matter
could be candidate for dark matter
\cite{Cheung:2007ut,Feldman:2007wj}.

Consider for specificity that the hidden sector contains Dirac fermions.
Such fermions $(\chi_m)$ will couple to a $Z'$  with normal electroweak strength and thus
can produce a significant size decay width for $Z'$ into
ordinary quarks and leptons when $M_{\chi_m}$ is below $M_{Z'}/2$
and dark matter constraints can be easily satisfied\cite{Cheung:2007ut}.
However, one consequence of this phenomenon is that the dilepton signal associated
with the decay of the $Z'$  into ordinary leptons will be highly suppressed
because of the significantly larger decay of the $Z'$ into the hidden sector fermions.
Further,
it would at first  appear that the mechanism above for the satisfaction of relic density constraints
may not work when the Dirac fermion mass in above $M_{Z'}/2$.
However, it is well known that the thermal averaging over the poles  for annihilations
in the early universe  can allow one to satisfy the relic density constraints.
 The mechanism comes into play when the Dirac fermion mass $M_{\chi_m}$ is larger than
$M_{Z'}$ and indeed in this case it is possible to satisfy the WMAP constraints over a
significant part of the parameter space. Further, in this case one also has a strong
dileptonic signal for the $Z'$ which is accessible at the Tevatron and at the 
LHC\cite{Feldman:2007wj,Feldman:2006ce}.
\\\\
{\bf Concluding Remarks}: We summarize now our results.  We have shown that in
 a broad class of models one finds  the existence of
a Wall consisting of a copious number of parameter points  in the Chargino
Patterns.
The chances of discovery of dark matter on the Wall are enhanced due to
clustering.  The neutralino-proton scalar cross  sections at the Wall is
$\sigma_{SI}(\chi p)\sim 10^{-44\pm .5} {\rm cm}^2$
well within the reach of the next generation of dark matter experiments.
We have also argued that the direct detection of dark matter along with the LHC signatures
provide a dual probe of
SUSY. Thus in some cases dark matter detection can probe the parameter space of
supergravity models which may not be easily accessible at  the LHC at least with low
luminosity in multilepton modes.
Thus the direct detection of dark matter and LHC signatures are complementary
in their probe of  SUSY.
Finally, we have argued that the
hidden sector is a viable source of dark matter. Specifically we have discussed two cases
regarding dark matter from the hidden sector. These include an extra weakly interacting Majorana
dark  matter candidate which is a linear combination of fields in the hidden sector and the
connector sector, and a milli-charged dark matter matter candidate which arises from matter
Dirac fermions  in the hidden sector.
\\

This research is supported in part by the U.S.
NSF Grant No. NSF-PHY-0757959.

\vspace{-.5cm}    

%
%


\end{document}